\documentclass[%
11pt,
superscriptaddress,
preprintnumbers,
tightenlines,
showpacs,
showkeys,
nofootinbib,
amsmath,
amssymb,
aps,
prd,
floatfix,
]{revtex4-1}

\usepackage{graphicx}

\newcommand{\be}{\begin{equation}}
\newcommand{\ee}{\end{equation}}
\newcommand{\bea}{\begin{eqnarray}}
\newcommand{\eea}{\end{eqnarray}}
\newcommand{\non}{\nonumber}
\newcommand{\bi}{\begin{itemize}}
\newcommand{\ei}{\end{itemize}}

\begin{document}

\title
{
Curvature of the critical line on the plane of quark chemical potential and pseudo scalar meson mass for three-flavor QCD
} 

\author{Xiao-Yong Jin\footnote{Present address: 
Argonne Leadership Computing Facility, Argonne National Laboratory, Argonne, IL 60439, USA. }}
\affiliation{RIKEN Advanced Institute for Computational Science, Kobe, Hyogo 650-0047, Japan}

\author{Yoshinobu Kuramashi}
\affiliation{Faculty of Pure and Applied Sciences, University of Tsukuba, Tsukuba, Ibaraki 305-8571, Japan}
\affiliation{Center for Computational Sciences, University of Tsukuba, Tsukuba, Ibaraki 305-8577, Japan}
\affiliation{RIKEN Advanced Institute for Computational Science, Kobe, Hyogo 650-0047, Japan}

\author{Yoshifumi~Nakamura}
\affiliation{RIKEN Advanced Institute for Computational Science, Kobe, Hyogo 650-0047, Japan}
\affiliation{Graduate School of System Informatics, Department of Computational Sciences, Kobe University, Kobe, Hyogo 657-8501, Japan}

\author{Shinji Takeda}\email[]{takeda@hep.s.kanazawa-u.ac.jp}
\affiliation{Institute of Physics, Kanazawa University, Kanazawa 920-1192, Japan}
\affiliation{RIKEN Advanced Institute for Computational Science, Kobe, Hyogo 650-0047, Japan}

\author{Akira Ukawa}
\affiliation{RIKEN Advanced Institute for Computational Science, Kobe, Hyogo 650-0047, Japan}

\date{\today}
\begin{abstract}
We investigate the phase structure of three-flavor QCD in the presence of finite quark chemical potential $\mu/T\lesssim1.2$ by using the non-perturbatively $O(a)$ improved Wilson fermion action on lattices with a fixed temporal extent $N_{\rm t}=6$ and varied spatial linear extents $N_{\rm s}=8,10,12$.
Especially, we focus on locating the critical end point that characterizes the phase structure, and extracting the curvature of the critical line on the $\mu$-$m_{\pi}$ plane.
For Wilson-type fermions, the correspondence between bare parameters and physical parameters is indirect. 
Hence we present a strategy to transfer the bare parameter phase structure to the physical one, in order to obtain the curvature.
Our conclusion is that the curvature is positive.  This implies that, if one starts from a quark mass in the region of crossover at zero chemical potential, one would encounter a first-order phase transition when one raises the chemical potential.  
\end{abstract}

\preprint{UTHEP-670, UTCCS-P-80, KANAZAWA-15-01}

\pacs{11.15.Ha,12.38.Gc,25.75.Nq}

\maketitle


\section{Introduction}
\label{sec:int}

At zero baryon number density, on the two-dimensional plane spanned by the light (up-down degenerated) quark mass $m_{ud}$ and strange quark mass $m_{s}$, the first order phase transition around the massless point $m_{ud}=m_s=0$ becomes weaker as the quark masses increase, 
and eventually turns into a crossover at some  finite quark masses.
The boundary between the first order phase transition region and the crossover region forms a line of second order phase transition,  called the critical end line.

A question of obvious importance is the location of the critical line.  
Monte Carlo results on this issue are rather confusing at present.
For the staggered fermion action, recent studies with improved action could place only an upper bound on the three-flavor degenerate  critical quark mass, $m$, which is very small in the range of $m/m_{ud}^{phys}\approx 0.1$~\cite{Endrodi:2007gc,Ding:2011du}.  
This is in contrast to recent as well as an earlier study  with the naive action~\cite{Kayaetal1999,Karsch:2001nf,deForcrand:2006pv,Smith11} which observed first order signals up to $m/m_{ud}^{phys}\approx 2-3$.   
Furthermore, our recent study with the Wilson-clover fermion action~\cite{Jin:2014hea}, motivated in part by the unclear status with the staggered action,  could identify the critical end point, although the cut-off dependence of the location is rather large.

The location of the critical end point in the QCD phase diagram with finite density is also an important issue.
The first serious study on critical end point in QCD was given by Fodor and Katz who employ Lee-Yang zero analysis \cite{Fodor:2001au,Fodor:2001pe}.
After this study, various attempts were made and here we quote some reviews \cite{deForcrand:2010ys,Philipsen:2011zx} about such studies.
In this article, we address an issue of how the critical end line extends when switching on the chemical potential.
An interesting result was reported in \cite{deForcrand:2006pv,deForcrand:2007rq} which explored the imaginary chemical potential approach with the naive staggered fermion action. 
There it was observed that the critical surface has a negative curvature in the $\mu$ direction. This means that a first-order phase transition at zero chemical potential disappears when the chemical potential is increased, rather contrary to one's naive guess.  
Our purpose in this paper is to study this question by simulations with real chemical potential using the Wilson-clover fermion action.  This is a natural sequel of our work in \cite{Jin:2014hea}.

The rest of the paper is organized as follows.
In section \ref{sec:strategy}, we explain a strategy on how to draw the critical line on the $\mu$-$m_{\pi}$ plane.
Simulation details including the parameters and the simulation algorithm are summarized in section \ref{sec:details}.
We present numerical results in section \ref{sec:results}.
Finally, concluding remarks are given in section \ref{sec:conclusion}.

\section{Strategy}
\label{sec:strategy}
Let us explain our strategy to survey the phase space for $N_{\rm f}=3$ QCD in order to  identify the critical end point for the Wilson-type fermions.
The final goal of this section is to show how to obtain the curvature of the critical end line on the $\mu$-$m_{\pi}$ plane.
Note that in this section we do not use lattice units when expressing dimensionful physical quantities.

\begin{figure}[t]
\begin{center}
\begin{tabular}{cc}
\hspace{-12mm}
\scalebox{0.4}{\includegraphics{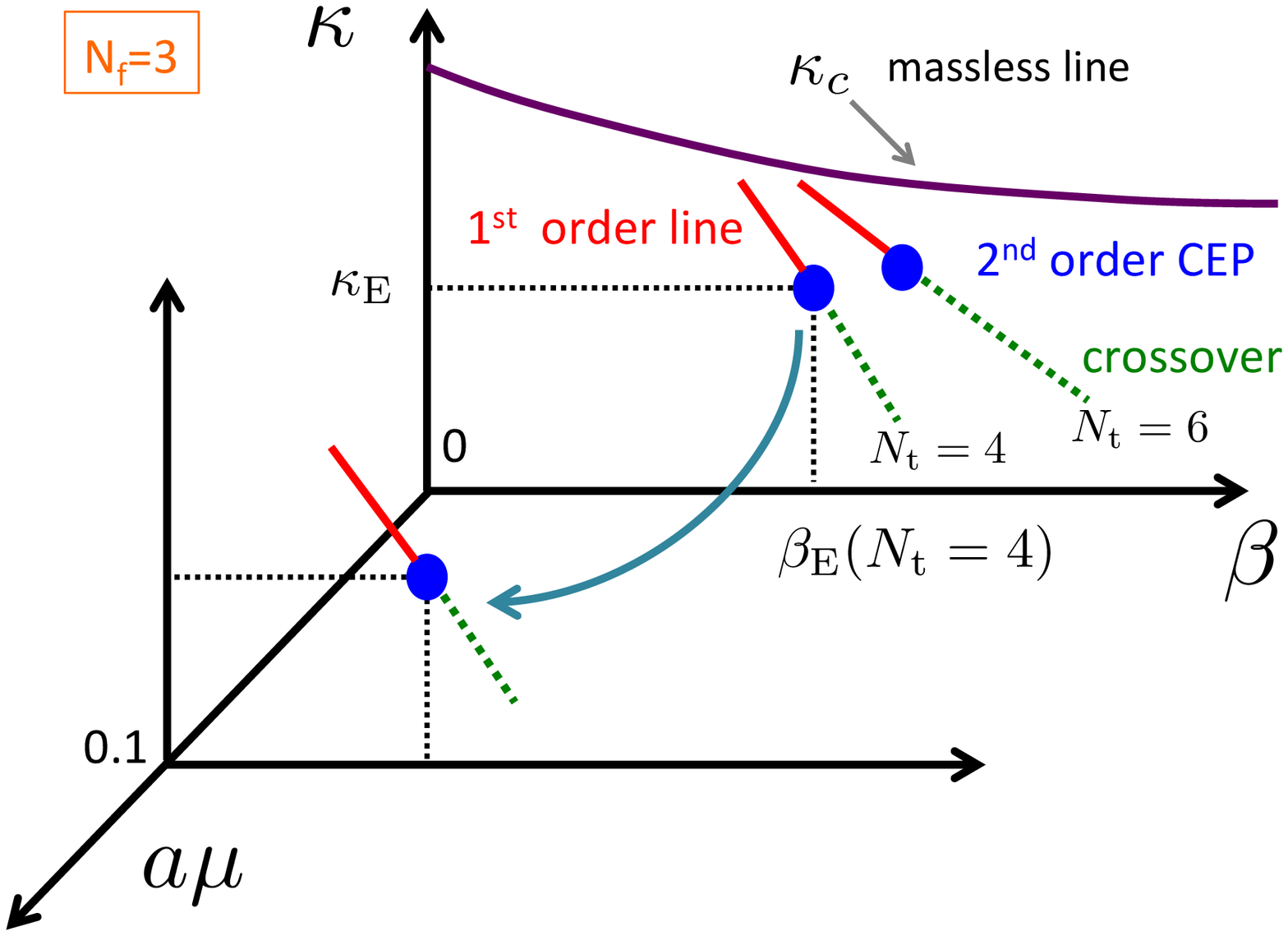}}
&
\hspace{-13mm}
\scalebox{0.4}{\includegraphics{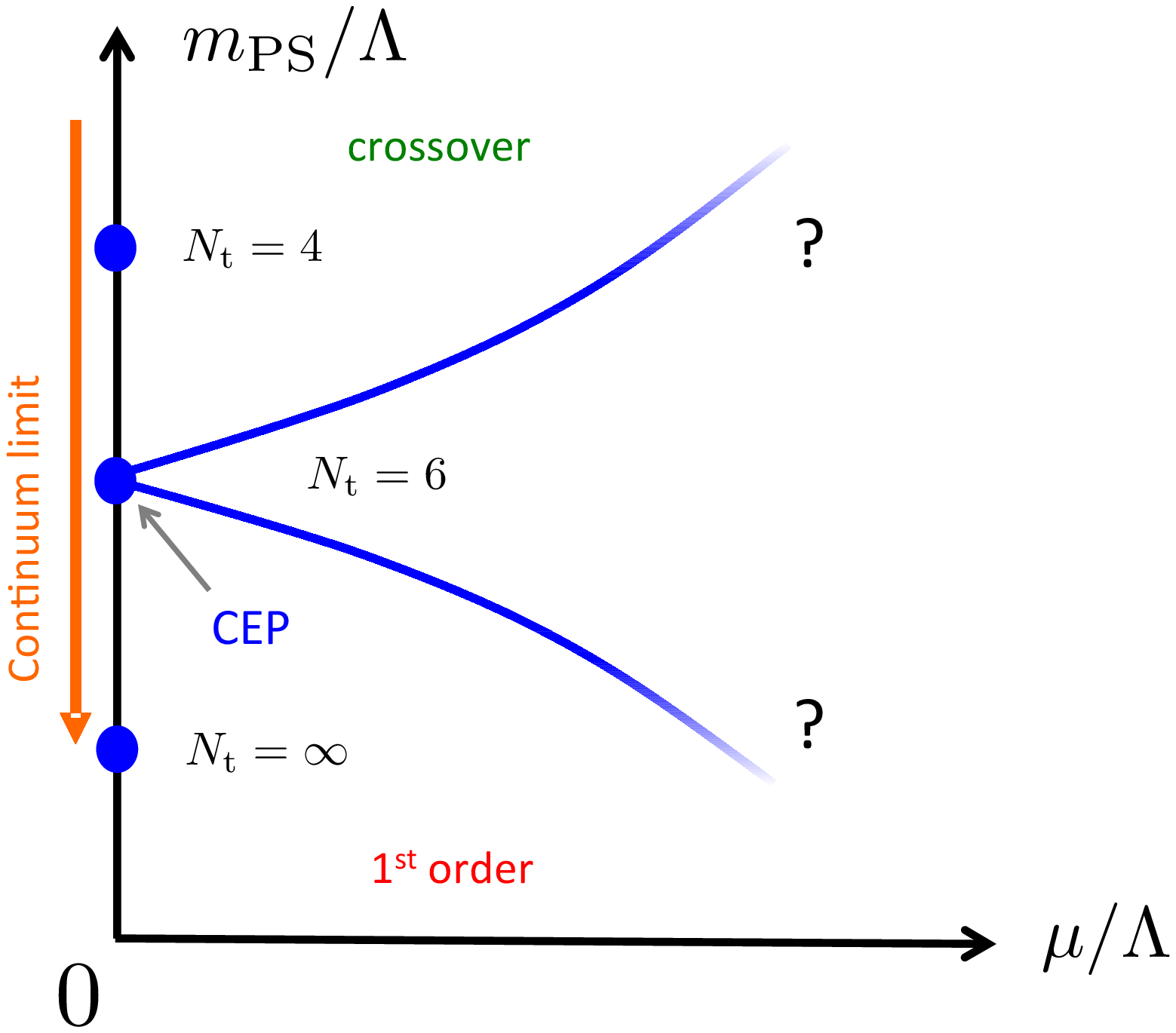}}
\\
\end{tabular}
\end{center}
\caption{
Strategy:
The left panel is the phase diagram for bare parameters spanned by $\beta$, $\kappa$ and $a\mu$ for $N_{\rm f}=3$.
The right panel is the same phase diagram but depicted for physical parameters spanned by $m_{\rm PS}/\Lambda$ and $\mu/\Lambda$ where $\Lambda$ is some reference physical quantity at zero density.
The blue line extending from the critical end point at $a\mu=0$ is the critical line.
We study the signature of the curvature of the critical line with fixed $N_{\rm t}=6$.
}
\label{fig:strategy}
\end{figure}

First we consider the zero density case.
Since the quark masses are all degenerate, we have only two bare parameters $\beta$ and $\kappa$ ($a\mu=0$ plane in the left panel in Figure~\ref{fig:strategy}).
For a given temporal lattice size, say $N_{\rm t}=4$, by using the peak position of susceptibility or zero of skewness of quark condensate, one can draw the line of  finite temperature transition (the solid red line and the dotted green line in the left panel in Figure~\ref{fig:strategy}).
The transition changes from being of first order to cross over at a second order critical end point (the blue point in the left panel in Figure~\ref{fig:strategy}). 
We compute the kurtosis (which is the Binder cumulant minus three) of quark condensate along the transition line for a set of spatial volumes $N_{\rm s}^3$.
The intersection point is identified as the critical end point \cite{Karsch:2001nf}.
In this way, we can determine the critical end point in the bare parameter space
$(\beta_{\rm E},\kappa_{\rm E})$ and this procedure can be repeated for other values of $N_{\rm t}$.

In order to translate the critical end point in the bare parameter space to that in the physical parameter space, we measure dimensionless ratios of pseudo-scalar meson mass and some reference quantity with mass-dimension one $m_{\rm PS}/\Lambda$ for the bare parameters $(\beta_{\rm E},\kappa_{\rm E})$ by a zero temperature simulation.
One can choose any reference quantity $\Lambda$, say $T$ (temperature), $1/\sqrt{t_0}$ (Wilson flow) \cite{wilflow} or $m_{\rm V}$ (vector meson mass).
To avoid the multiplicative renormalization issue, we  use $m_{\rm PS}$ in the numerator of the ratio and not quark masses.
In this way we pin down the critical end point (the blue point in the right panel in Figure~\ref{fig:strategy}) in the physical parameter space whose axes are given by $m_{\rm PS}/\Lambda$ and $\mu/\Lambda$. 
By repeating the same calculation for increasingly larger values of $N_{\rm t}$, we can take the continuum limit (the orange downward arrow in the right panel in Figure~\ref{fig:strategy}) of the critical end point in the physical parameter space at zero density,
\begin{equation}
\frac{m^{\rm cont}_{\rm PS,E}(\mu=0)}{\Lambda^{\rm cont}_{\rm E}(\mu=0)}
=
\lim_{N_{\rm t}\rightarrow\infty}
\frac{m_{\rm PS,E}(\mu=0)}{\Lambda_{\rm E}(\mu=0)}.
\end{equation}
This strategy is in fact used in our zero density study \cite{Jin:2014hea}.

When switching on the chemical potential, the basic procedure is the same;  one just has to repeat the same analysis on a different plane with $\mu\neq0$ (See the left panel in Figure~\ref{fig:strategy}).
For a fixed lattice temporal size, $N_{\rm t}=6$,
in order to draw the critical end line, we consider a pair of dimensionless ratios
\begin{equation}
\frac{m_{\rm PS,E}(\mu)}{m_{\rm PS,E}(0)}
\hspace{10mm}
\mbox{and}
\hspace{10mm}
\frac{\mu}{T_{\rm E}(0)},
\label{eqn:ratios}
\end{equation}
where for each ratio we have chosen proper reference quantities at zero density.
By plotting these two quantities one can obtain a critical line as shown in the right panel of Figure~\ref{fig:strategy}.
We are interested in seeing whether the critical line bends toward the lighter mass or heavier mass direction.
More quantitatively, from a fitting
\begin{equation}
\left(
\frac{m_{\rm PS,E}(\mu)}{m_{\rm PS,E}(0)}
\right)^2
=
1
+
\alpha_1
\left(
\frac{\mu}{\pi T_{\rm E}(0)}
\right)^2
+
\alpha_2
\left(
\frac{\mu}{\pi T_{\rm E}(0)}
\right)^4
+...,
\label{eqn:MPSmuT}
\end{equation}
we shall extract the curvature $\alpha_1$ and see its sign, and this is the final goal of this paper.

If one wants to take the continuum limit of the critical end line, one has to take the $N_{\rm t}\rightarrow\infty$ limit for fixed values of $\mu/T_{\rm E}(0)$ 
\begin{equation}
\frac{m^{\rm cont}_{\rm PS,E}(\mu)}{m^{\rm cont}_{\rm PS,E}(0)}
=
\left.
\lim_{N_{\rm t}\rightarrow\infty}
\frac{m_{\rm PS,E}(\mu)}{m_{\rm PS,E}(0)}
\right|_{ \mbox{ fixed } \mu/T_{\rm E}(0)}.
\end{equation}
After repeating the same procedure with different values of $\mu/T_{\rm E}(0)$, one can plot
$(m^{\rm cont}_{\rm PS,E}(\mu)/m^{\rm cont}_{\rm PS,E}(0))^2$
as a function of
$\mu/T_{\rm E}(0)$.
Then by fitting with the same form as in eq.(\ref{eqn:MPSmuT}),
one can obtain the curvature in the continuum limit.

\section{Simulation details}
\label{sec:details}

We employ the Wilson-clover fermion action with non-perturbatively tuned $c_{\rm sw}$~\cite{CPPACS2006} in the presence of chemical potential with the anti-periodic boundary condition in the temporal direction for fermion fields while the periodic boundary condition is imposed for spatial direction.
The Iwasaki gauge action \cite{iwasaki} is used for the gluon sector and gauge link variables satisfy the periodic boundary condition.
The number of flavor is three, $N_{\rm f}=3$, and the masses and chemical potentials for quarks are all degenerate.
The temporal lattice size and the simulated quark chemical potential are fixed to $N_{\rm t}=6$ and $a\mu=0.1$, respectively, and thus $\mu/T=0.6$. 
In our study, the phase reweighting method explained below is used to deal with the complex phase, and
to survey a wide range of $\mu$ and $\kappa$, we adopt the multi-parameter reweighting method; details are given in Appendix \ref{sec:reweighting}.
To perform  finite size scaling analysis, the spatial volume is changed over the linear sizes $N_{\rm s}=8$, $10$ and $12$.
In order to search for the transition point, we select four $\beta$ points ($\beta=1.70$, $1.73$, $1.75$ and $1.77$) and for each $\beta$, we vary $\kappa$ to locate the transition point.

The phase reweighting method is adopted to handle the complex phase according to 
\begin{equation}
\langle
{\cal O}
\rangle
=
\frac{
\langle
{\cal O} e^{iN_{\rm f}\theta}
\rangle_{||}
}
{
\langle
e^{iN_{\rm f}\theta}
\rangle_{||}
},
\end{equation}
where $
\langle
...
\rangle_{||}
$ is the average with phase quenched fermion determinant
\begin{equation}
{\cal Z}_{||}
=
\int [dU]
e^{-S_{\rm G}}
|\det D(\mu)|^{N_{\rm f}},
\end{equation}
and the phase factor for one-flavor is given by
\begin{equation}
e^{i\theta}
=
\frac{\det D(\mu)}{|\det D(\mu)|}.
\end{equation}
Configurations are generated by RHMC \cite{RHMC} with the phase quenched quark determinant.
The MD step size is chosen such that a reasonable acceptance rate $\gtrsim80\%$ is retained.
For each lattice parameter set $(\beta,\kappa,N_{\rm t},N_{\rm s})$ we generate $O(100,000)$ trajectories and the configurations are stored at every 10th trajectory;  the order of number of configurations are $O(10,000)$ for each parameter set.
The phase factor and $\mu$-derivatives of the fermion determinants required in $\mu$-parameter reweighting are computed exactly using the analytical reduction technique \cite{Danzer:2008xs,Takeda:2011vd,takedanote} for all stored configurations.
The dense matrix obtained by the reduction is numerically computed on GPGPU with LAPACK routines.
We measure the trace of quark propagator and its higher power up to fourth order which are used not only for the computation of higher moments of quark condensate but also for the parameter reweighting (See Appendix \ref{sec:reweighting} for details).
In the computation of traces, we adopt the noise method with 20 Gaussian noises that is checked to be sufficient to control the noise error.

For each fixed parameter set ($\beta$, $a\mu$, $N_{\rm t}$, $N_{\rm s}$), we make runs at several values of $\kappa$.
In order to integrate those runs we adopt the multi-ensemble reweighting technique~\cite{reweighting} and search for the transition point in $\kappa$ for the fixed parameter set.
See Appendix \ref{sec:multiensemblereweighting} for the details of the multi-ensemble reweighting.
Here we only mention that we use some approximation to efficiently evaluate the quark determinant in the reweighting factor as well as observables at many reweighting points.

In our approach, there are practically two important issues: the overlap problem and the validity of approximation
made at calculating the ratio of quark determinant in the reweighting factor.
The issue of the overlap problem will be addressed in the next section.
The validity of the approximation is discussed in Appendix \ref{sec:reweighting} and the conclusion is that the approximation we made is safe in our parameter region.

The physical scale settings we use in this paper, for example the Wilson flow scale $\sqrt{t_0}$ \cite{wilflow} and the hadron mass, are taken from Ref.~\cite{Jin:2014hea}.

\section{Results}
\label{sec:results}

\subsection{Phase reweighting factor}
\label{subsec:phasereweightingfactor}

\begin{figure}[t!]
\begin{center}
\begin{tabular}{c}
\scalebox{0.9}{\includegraphics{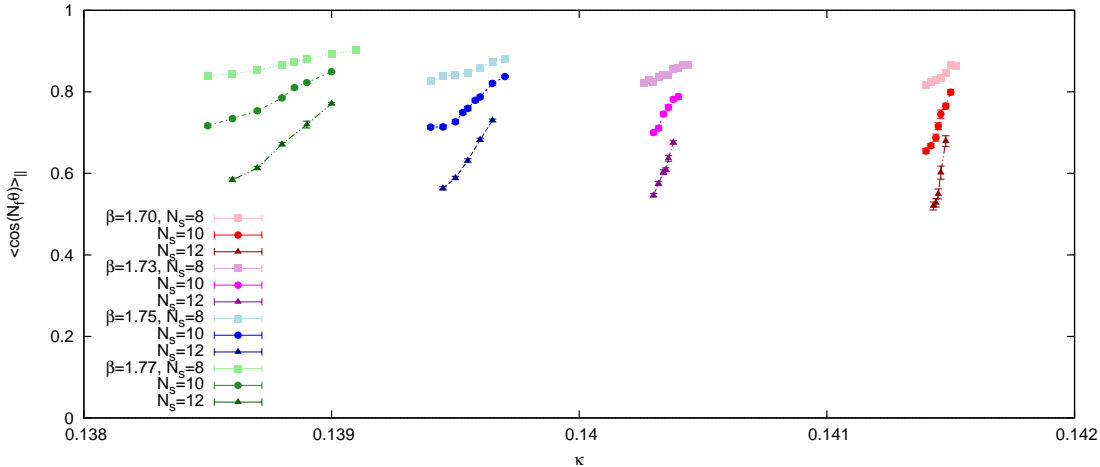}}
\end{tabular}
\end{center}
\caption{The average of phase reweighting factor with $N_{\rm f}=3$ as a function of $\kappa$.
$\mu/T=a\mu\times N_{\rm t}=0.1\times6=0.6$.
The reweighting factor is significantly away from zero.
This shows that the sign problem is mild in this region.
}
\label{fig:phasereweighting}
\end{figure}

Figure~\ref{fig:phasereweighting} shows the average value of the phase-reweighting factor as a function of $\kappa$.
For small $\kappa$ and large volumes, the value becomes smaller, 
signaling that the sign problem is becoming serious.
Nevertheless, it stays away from zero ($\gtrsim0.5$) beyond statistical error, guaranteeing the validity of 
the phase-reweighting for our range of lattice parameters.

\begin{figure}[t]
\begin{center}
\begin{tabular}{cc}
\scalebox{0.8}{\includegraphics{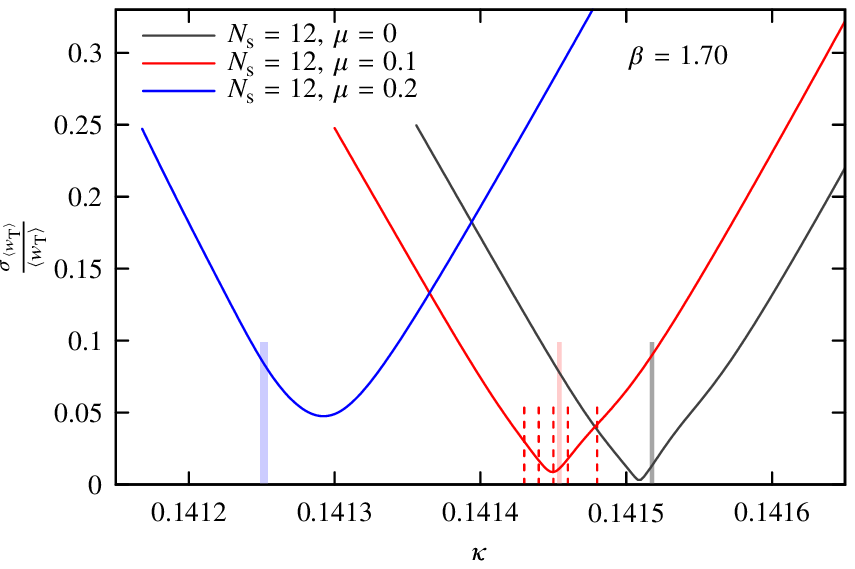}}
&
\scalebox{0.8}{\includegraphics{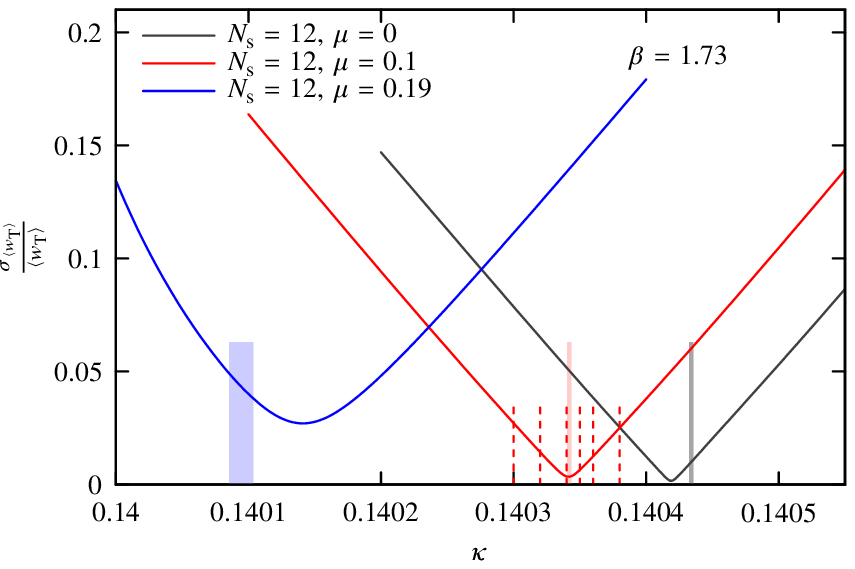}}
\\
\scalebox{0.8}{\includegraphics{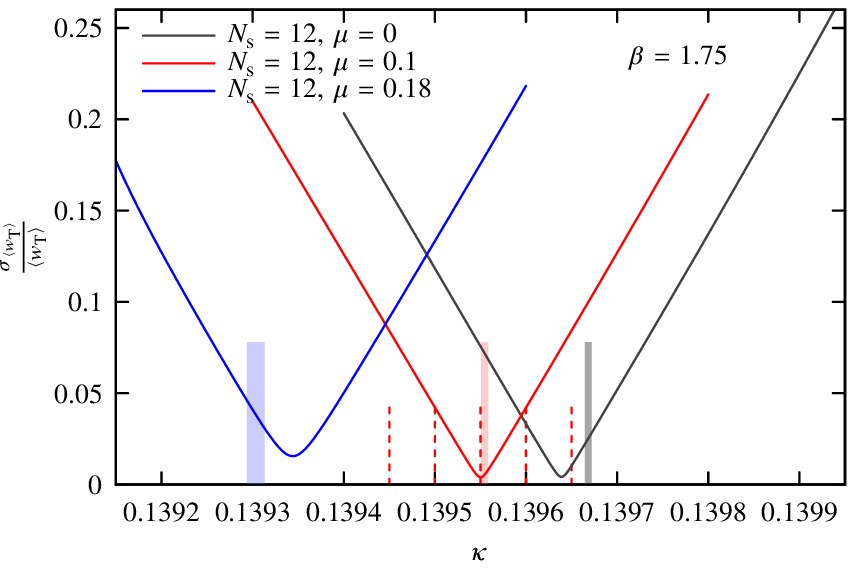}}
&
\scalebox{0.8}{\includegraphics{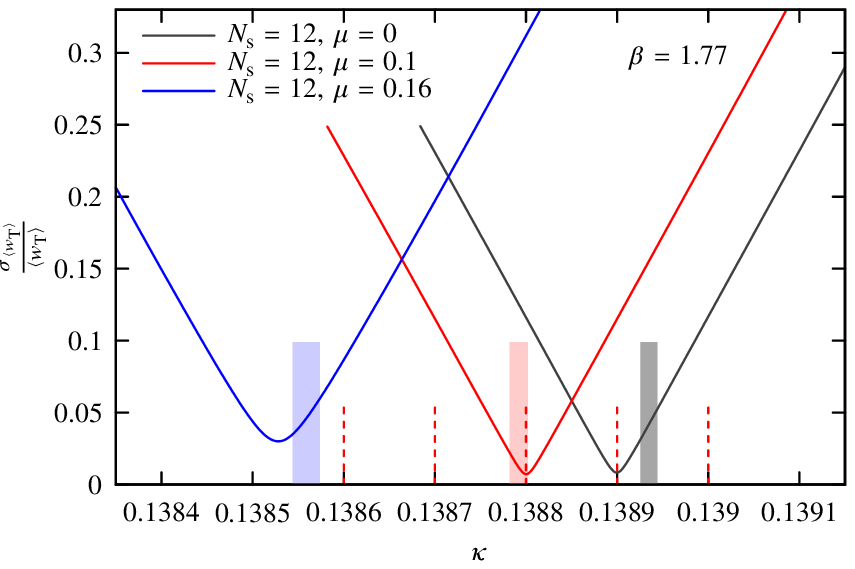}}
\end{tabular}
\end{center}
\caption{
Relative error of the reweighting factor of multi-ensemble reweighting in eq.(\ref{eqn:weight}) as a function of $\kappa$ for $\beta=1.70$ (left-top), $1.73$(right-top), $1.75$(bottom-left), $1.77$(bottom-right).
The shaded bands represent the respective $\kappa$ values at transition/cross-over, from Table~\ref{tab:transitionpoint} and the dotted lines indicate the location of the simulated $\kappa$ values.
Here only three selected values of the chemical potential are shown for each $\beta$.
The spatial lattice is fixed $N_{\rm s}=12$.
The relative weight is significantly smaller than one, thus
the sign problem is mild even for the reweighted parameter space.
}
\label{fig:reweightingfactor}
\end{figure}


We also check the average of the reweighting factor $w_{\rm T}$ in eq.(\ref{eqn:weight}) of the multi-ensemble reweighting, $\langle w_{\rm T}\rangle_{\rm ME}=\sum_{U} w_{\rm T}(U)/\sum_s N_s$ where unexplained notation is given in Appendix \ref{sec:multiensemblereweighting}. The relative error of the reweighting factor is plotted in Figure \ref{fig:reweightingfactor}.
The errors are estimated by the jackknife method with bin size of $1000$ configurations.
Figure~\ref{fig:reweightingfactor} shows that the relative error is sufficiently small $[\mbox{error of }\langle w_{\rm T}\rangle_{\rm ME}]/\langle w_{\rm T}\rangle_{\rm ME}\ll1$, even at larger chemical potential $a\mu\approx0.2$.
This means that the central value of the reweighting factor is significantly away from zero beyond many sigmas.
Thus, we conclude that the overlap problem is not so severe in our parameter region.

\subsection{Moments of chiral condensate and transition point}
\label{subsec:momentTP}
\begin{figure}[t]
\begin{center}
\begin{tabular}{cc}
\hspace{-10mm}
\scalebox{0.75}{\includegraphics{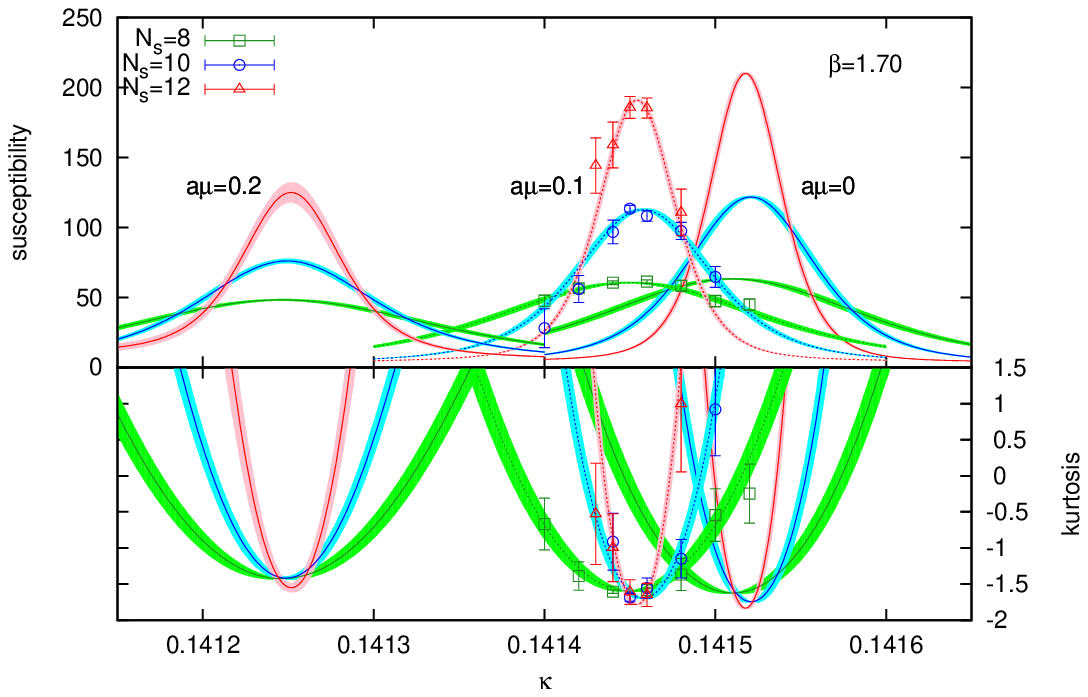}}
&
\hspace{-10mm}
\scalebox{0.75}{\includegraphics{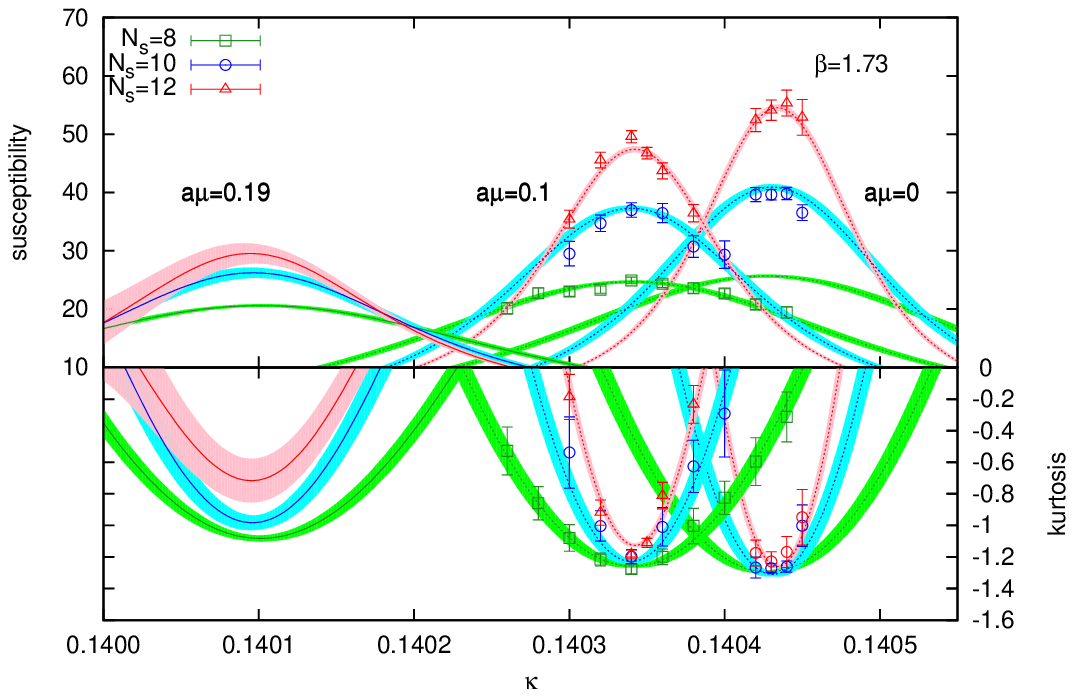}}
\\
\hspace{-10mm}
\scalebox{0.75}{\includegraphics{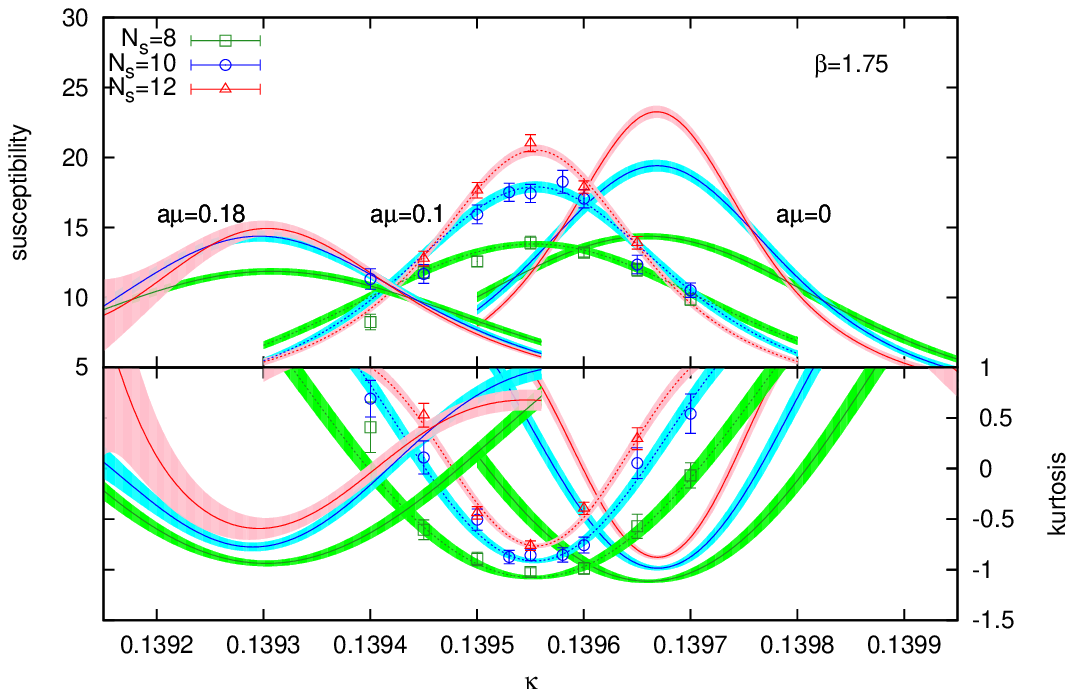}}
&
\hspace{-10mm}
\scalebox{0.75}{\includegraphics{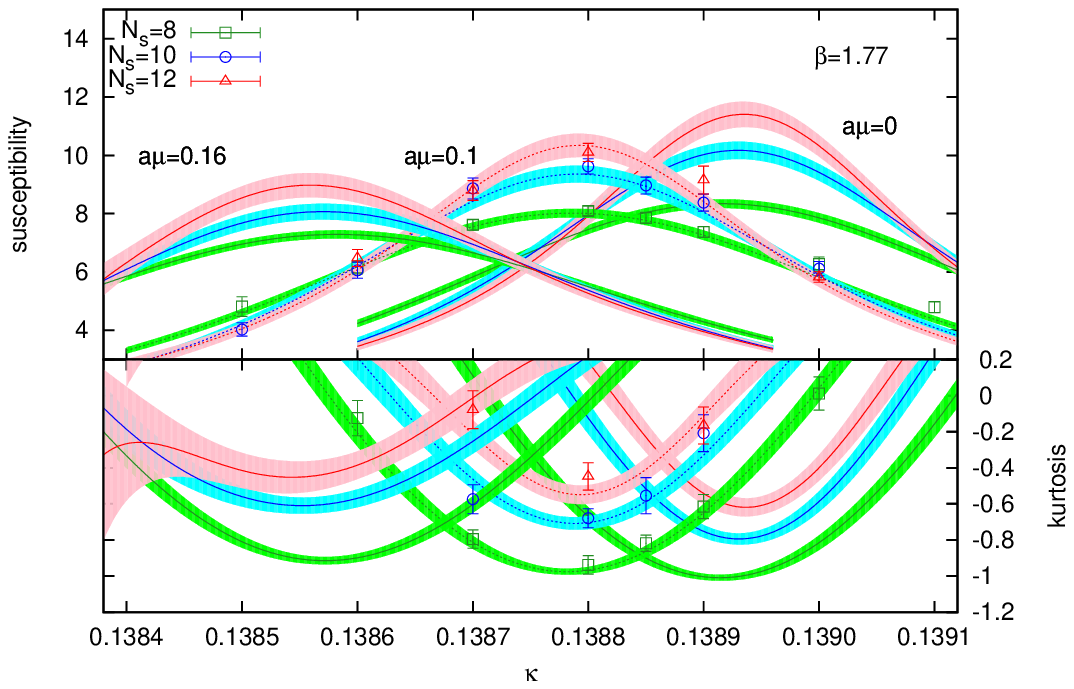}}
\end{tabular}
\end{center}
\caption{
The susceptibility and kurtosis of quark condensate as a function of $\kappa$ for 
$\beta=1.70$ (left-top), $1.73$(right-top), $1.75$(bottom-left), $1.77$(bottom-right).
In each figure, in addition to the raw data at $a\mu=0.1$ with $N_{\rm s}=8,10,12$, we plot the reweighting results expressed by a band with selected 3 values of $a\mu$.
All reweighting results are produced from configurations at $a\mu=0.1$ and several ensembles with different $\kappa$ values are integrated into the multi-ensemble reweighting.
In $\beta=1.73$ with $N_{\rm s}=10,12$, we also plot the raw data at $a\mu=0$ given in the zero density study \cite{Jin:2014hea}.
These raw data and the reweighting results are consistent with each other, although they are completely independent.
This shows that the reweighting together with the approximation used for the reweighting factors and observables (See Appendix \ref{sec:reweighting} for details of the approximation) is fine.
}
\label{fig:moments}
\end{figure}

{\tiny
\begin{table}[t]
\begin{tabular}{llr|lrl}
\hline
\hline
$\beta$
\phantom{aaa}
&
$a\mu$
\phantom{aaa}
&
$N_{\rm s}$
&
$\kappa_{\rm t}$
&
\phantom{aa}
$\chi_{\max}$
\phantom{aaaa}
&
$K_{\min}$
\\
\hline
$ 1.70 $&$ 0.00 $&$ 8 $&$ 0.1415100 ( 52 ) $&$ 63.25 ( 57 ) $&$ -1.6237 ( 72 ) $\\
$ 1.70 $&$ 0.00 $&$ 10 $&$ 0.1415209 ( 28 ) $&$ 121.74 ( 91 ) $&$ -1.7447 ( 50 ) $\\
$ 1.70 $&$ 0.00 $&$ 12 $&$ 0.1415177 ( 15 ) $&$ 210.09 ( 90 ) $&$ -1.8336 ( 24 ) $\\
$ 1.70 $&$ 0.00 $&$\infty$&$ 0.1415203 ( 27 ) $& & \\
\cline{2-6}
$ 1.70 $&$ 0.10 $&$ 8 $&$ 0.1414498 ( 52 ) $&$ 60.54 ( 57 ) $&$ -1.5946 ( 83 ) $\\
$ 1.70 $&$ 0.10 $&$ 10 $&$ 0.1414577 ( 28 ) $&$ 112.71 ( 93 ) $&$ -1.6951 ( 63 ) $\\
$ 1.70 $&$ 0.10 $&$ 12 $&$ 0.1414541 ( 15 ) $&$ 190.9 ( 1.0 ) $&$ -1.7826 ( 38 ) $\\
$ 1.70 $&$ 0.10 $&$\infty$&$ 0.1414554 ( 31 ) $& & \\
\cline{2-6}
$ 1.70 $&$ 0.20 $&$ 8 $&$ 0.1412465 ( 59 ) $&$ 48.22 ( 65 ) $&$ -1.424 ( 15 ) $\\
$ 1.70 $&$ 0.20 $&$ 10 $&$ 0.1412493 ( 32 ) $&$ 76.0 ( 1.7 ) $&$ -1.417 ( 25 ) $\\
$ 1.70 $&$ 0.20 $&$ 12 $&$ 0.1412516 ( 27 ) $&$ 125.0 ( 7.3 ) $&$ -1.550 ( 64 ) $\\
$ 1.70 $&$ 0.20 $&$\infty$&$ 0.1412537 ( 46 ) $& & \\
\hline
$ 1.73 $&$ 0.00 $&$ 8 $&$ 0.1404267 ( 44 ) $&$ 25.66 ( 20 ) $&$ -1.2943 ( 84 ) $\\
$ 1.73 $&$ 0.00 $&$ 10 $&$ 0.1404304 ( 37 ) $&$ 40.90 ( 51 ) $&$ -1.310 ( 12 ) $\\
$ 1.73 $&$ 0.00 $&$ 12 $&$ 0.1404340 ( 16 ) $&$ 54.54 ( 48 ) $&$ -1.2600 ( 94 ) $\\
$ 1.73 $&$ 0.00 $&$\infty$&$ 0.1404371 ( 29 ) $& & \\
\cline{2-6}
$ 1.73 $&$ 0.10 $&$ 8 $&$ 0.1403406 ( 44 ) $&$ 24.68 ( 20 ) $&$ -1.2580 ( 87 ) $\\
$ 1.73 $&$ 0.10 $&$ 10 $&$ 0.1403400 ( 38 ) $&$ 37.29 ( 47 ) $&$ -1.228 ( 13 ) $\\
$ 1.73 $&$ 0.10 $&$ 12 $&$ 0.1403421 ( 16 ) $&$ 47.41 ( 44 ) $&$ -1.126 ( 11 ) $\\
$ 1.73 $&$ 0.10 $&$\infty$&$ 0.1403427 ( 30 ) $& & \\
\cline{2-6}
$ 1.73 $&$ 0.19 $&$ 8 $&$ 0.1401016 ( 50 ) $&$ 20.60 ( 22 ) $&$ -1.082 ( 19 ) $\\
$ 1.73 $&$ 0.19 $&$ 10 $&$ 0.1400964 ( 59 ) $&$ 26.22 ( 82 ) $&$ -0.983 ( 53 ) $\\
$ 1.73 $&$ 0.19 $&$ 12 $&$ 0.1400946 ( 92 ) $&$ 29.5 ( 1.8 ) $&$ -0.72 ( 13 ) $\\
$ 1.73 $&$ 0.19 $&$\infty$&$ 0.1400913 ( 97 ) $& & \\
\hline
$ 1.75 $&$ 0.00 $&$ 8 $&$ 0.1396591 ( 87 ) $&$ 14.37 ( 21 ) $&$ -1.115 ( 16 ) $\\
$ 1.75 $&$ 0.00 $&$ 10 $&$ 0.1396684 ( 58 ) $&$ 19.41 ( 41 ) $&$ -0.985 ( 21 ) $\\
$ 1.75 $&$ 0.00 $&$ 12 $&$ 0.1396682 ( 37 ) $&$ 23.27 ( 42 ) $&$ -0.878 ( 20 ) $\\
$ 1.75 $&$ 0.00 $&$\infty$&$ 0.1396722 ( 64 ) $& & \\
\cline{2-6}
$ 1.75 $&$ 0.10 $&$ 8 $&$ 0.1395533 ( 87 ) $&$ 13.82 ( 21 ) $&$ -1.077 ( 17 ) $\\
$ 1.75 $&$ 0.10 $&$ 10 $&$ 0.1395547 ( 59 ) $&$ 17.90 ( 39 ) $&$ -0.914 ( 21 ) $\\
$ 1.75 $&$ 0.10 $&$ 12 $&$ 0.1395546 ( 39 ) $&$ 20.52 ( 38 ) $&$ -0.765 ( 21 ) $\\
$ 1.75 $&$ 0.10 $&$\infty$&$ 0.1395552 ( 67 ) $& & \\
\cline{2-6}
\cline{2-6}
$ 1.75 $&$ 0.18 $&$ 8 $&$ 0.1393077 ( 97 ) $&$ 11.86 ( 22 ) $&$ -0.938 ( 28 ) $\\
$ 1.75 $&$ 0.18 $&$ 10 $&$ 0.1392962 ( 75 ) $&$ 14.37 ( 38 ) $&$ -0.776 ( 42 ) $\\
$ 1.75 $&$ 0.18 $&$ 12 $&$ 0.1393035 ( 97 ) $&$ 14.93 ( 53 ) $&$ -0.59 ( 10 ) $\\
$ 1.75 $&$ 0.18 $&$\infty$&$ 0.139295 ( 12 ) $& & \\
\hline
$ 1.77 $&$ 0.00 $&$ 8 $&$ 0.1389157 ( 85 ) $&$ 8.34 ( 15 ) $&$ -1.009 ( 16 ) $\\
$ 1.77 $&$ 0.00 $&$ 10 $&$ 0.1389300 ( 88 ) $&$ 10.17 ( 29 ) $&$ -0.794 ( 31 ) $\\
$ 1.77 $&$ 0.00 $&$ 12 $&$ 0.1389349 ( 93 ) $&$ 11.41 ( 43 ) $&$ -0.619 ( 50 ) $\\
$ 1.77 $&$ 0.00 $&$\infty$&$ 0.138944 ( 11 ) $& & \\
\cline{2-6}
$ 1.77 $&$ 0.10 $&$ 8 $&$ 0.1387866 ( 88 ) $&$ 8.02 ( 14 ) $&$ -0.976 ( 16 ) $\\
$ 1.77 $&$ 0.10 $&$ 10 $&$ 0.1387902 ( 98 ) $&$ 9.36 ( 29 ) $&$ -0.708 ( 34 ) $\\
$ 1.77 $&$ 0.10 $&$ 12 $&$ 0.138792 ( 10 ) $&$ 10.34 ( 40 ) $&$ -0.549 ( 51 ) $\\
$ 1.77 $&$ 0.10 $&$\infty$&$ 0.138794 ( 13 ) $& & \\
\cline{2-6}
$ 1.77 $&$ 0.16 $&$ 8 $&$ 0.1385825 ( 95 ) $&$ 7.29 ( 14 ) $&$ -0.914 ( 20 ) $\\
$ 1.77 $&$ 0.16 $&$ 10 $&$ 0.138570 ( 13 ) $&$ 8.07 ( 27 ) $&$ -0.610 ( 45 ) $\\
$ 1.77 $&$ 0.16 $&$ 12 $&$ 0.138559 ( 15 ) $&$ 8.98 ( 41 ) $&$ -0.453 ( 88 ) $\\
$ 1.77 $&$ 0.16 $&$\infty$&$ 0.138553 ( 18 ) $& & \\
\hline
\hline
\end{tabular}
\caption{
The transition point $\kappa_{\rm t}$, the peak hight of susceptibility and the minimum of kurtosis.
The errors are estimated by the jackknife analysis except for the value of $\kappa_{\rm t}$ at $N_{\rm s}=\infty$ where the error is calculated from the fit in eq.(\ref{eqn:kappa_t_fit}).
}
\label{tab:transitionpoint}
\end{table}
}

\begin{figure}[t]
\begin{center}
\begin{tabular}{c}
\scalebox{0.9}{\includegraphics{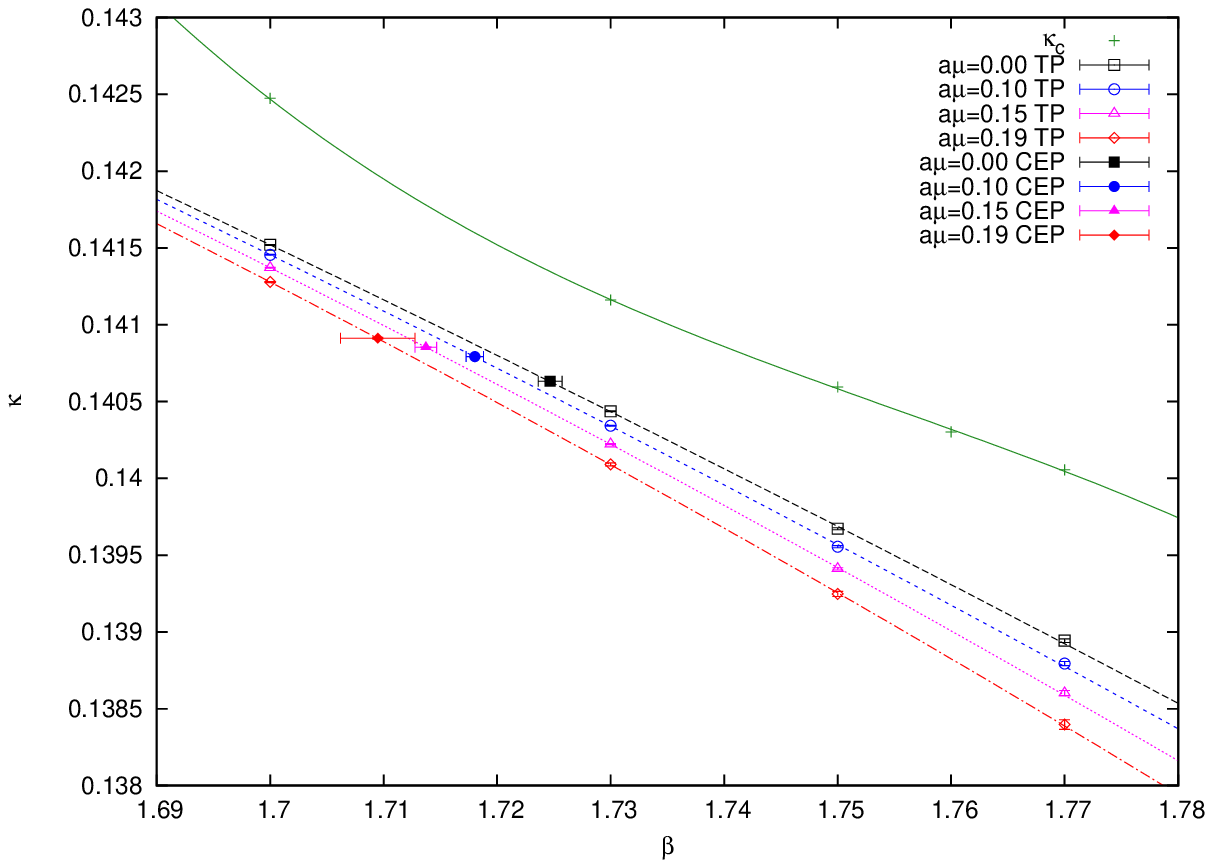}}
\end{tabular}
\end{center}
\caption{
The phase diagram of $N_{\rm f}=3$ QCD with finite chemical potential projected on the $(\beta,\kappa)$ plane.
The transition points are expressed by open symbols while the critical end points are given by filled ones.
The lower $\beta$ side of the critical end point is the first order phase transition region. 
For larger chemical potential, the critical point moves toward the upper-left corner.
The $\kappa_{\rm c}$ line where the pion mass vanishes is also shown.
}
\label{fig:betakappa}
\end{figure}

At finite quark mass, the quark bilinear operator $\overline{\psi}\psi$ is not a real order parameter but considered to be a mixture of ``energy" and ``magnetization" operators \cite{Karsch:2001nf}.
We study the bilinear operator as a primarily magnetization operator, however,  since we do not have enough data set to resolve the mixing of observables.
The detailed practical definition of its moments is given in Ref.~\cite{Jin:2014hea}.

Figure~\ref{fig:moments} shows curves of the susceptibility and kurtosis for quark condensate obtained by the multi-ensemble reweighting. 
The error bands are estimated by the jackknife method with bin size of $500-1000$ configurations.
For $a\mu=0.1$, the averages at each point of data generation are shown in order to illustrate how multi-ensemble curves interpolate those raw data. 
At $\beta=1.73$, the curves reweighted to $a\mu=0$ can be compared with data generated at zero density \cite{Jin:2014hea}.  The agreement supports the validity of multi-ensemble reweighting and jackknife error estimation away from $a\mu=0.1$. 
The applicable range of $\mu/\kappa$-reweighting depends on $\beta$, and judged from the growth of error, the lower $\beta$ tends to have a larger applicable range. 

As seen in the figures, the locations of the maximum of susceptibility and minimum of kurtosis are consistent with each other.
Furthermore the skewness zero location is also consistent with them although it is not shown here.
We take the location of the maximum of susceptibility as the transition point.  The numerical values are summarized in Table~\ref{tab:transitionpoint} where the peak height of susceptibility $\chi_{\rm max}$ and the minimum of kurtosis $K_{\rm min}$ are also listed for selected values of $a\mu$.

As seen in Table~\ref{tab:transitionpoint}, the volume dependence of the transition points is rather mild. Hence the thermodynamic limit can be safely taken with a fitting ansatz,
\begin{equation}
\kappa_{\rm t}(N_{\rm s})=\kappa_{\rm t}(\infty)+c/N_{\rm s}^3.
\label{eqn:kappa_t_fit}
\end{equation}
The resulting value of $\kappa_{\rm t}(\infty)$ is shown in Table~\ref{tab:transitionpoint}.
The phase diagram of bare parameters $\beta$ and $\kappa$ is given in Figure~\ref{fig:betakappa}.
The transition lines have a sensitivity on the value of chemical potential, $a\mu$.

\subsection{Kurtosis intersection}
\label{subsec:kurtosis intersection}

\begin{figure}[t]
\begin{center}
\begin{tabular}{cc}
\scalebox{1.}{\includegraphics{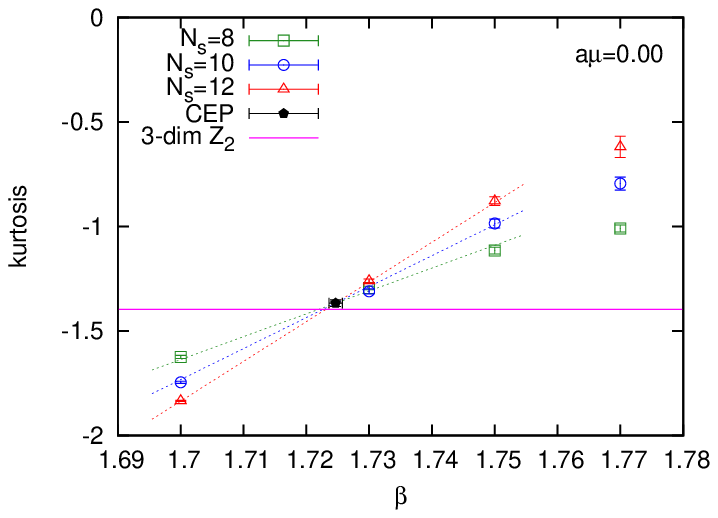}}
&
\scalebox{1.}{\includegraphics{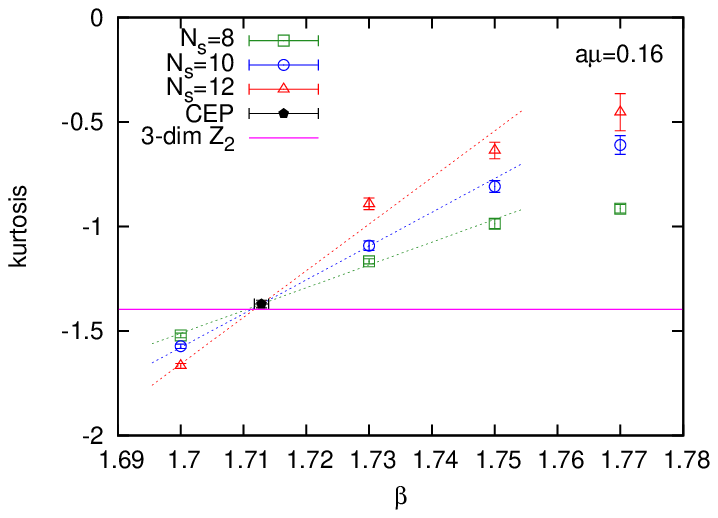}}
\\
\scalebox{1.}{\includegraphics{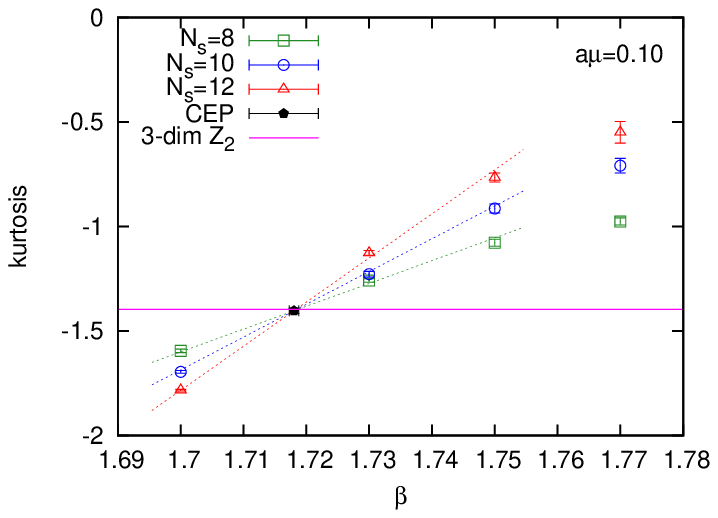}}
&
\scalebox{1.}{\includegraphics{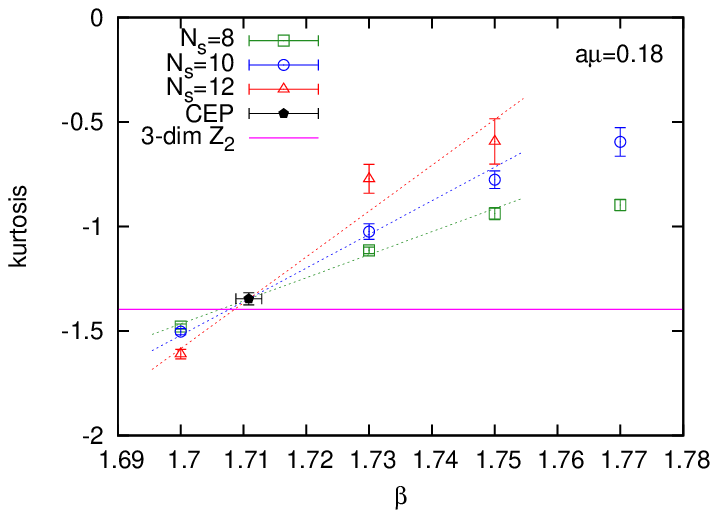}}
\\
\scalebox{1.}{\includegraphics{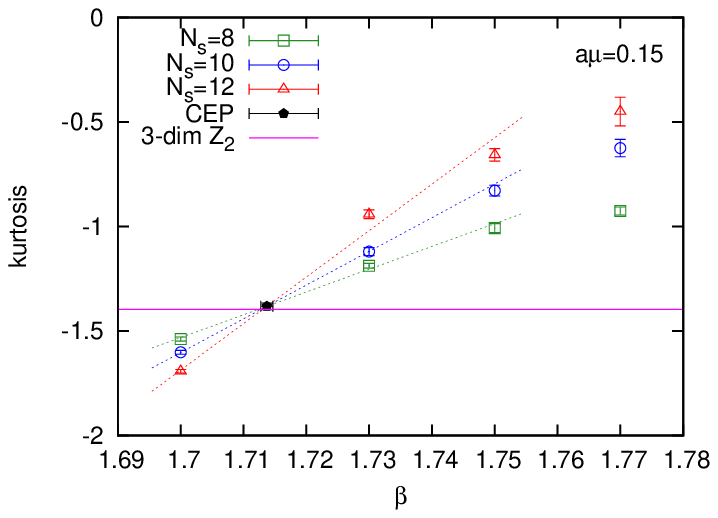}}
&
\scalebox{1.}{\includegraphics{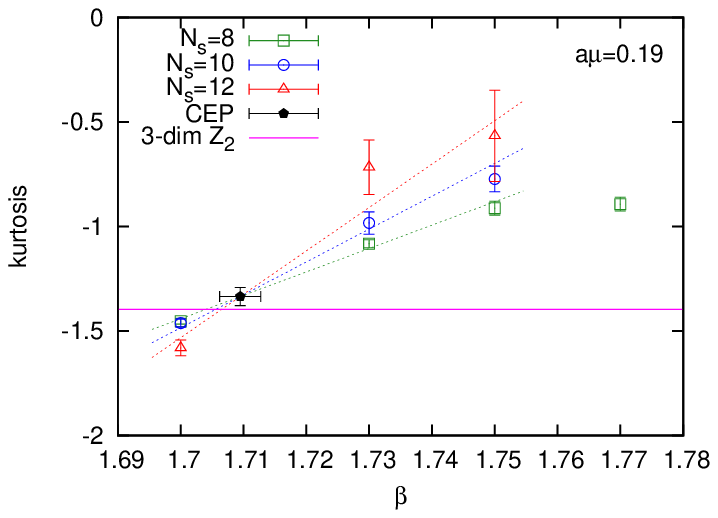}}
\end{tabular}
\end{center}
\caption{
Kurtosis intersection at $a\mu=0.00-0.19$.
In the fitting, three lowest values of $\beta$ are used.
The black pentagon represents the critical end point (CEP) in bare parameter space $\beta_{\rm E}$ and it moves to the lower side for larger chemical potential.
See Table \ref{tab:kurtosisintersectionfit} for the values of fitting parameters.
The horizontal magenta line shows $K_{\rm E}=-1.396$ for 3-dimensional ${\rm Z}_2$ universality class.
In this region of the chemical potential, the value of $K_{\rm E}$ is constant, namely the universality class does not change.
}
\label{fig:krtintersection}
\end{figure}

The next step is to determine the critical end point.
For that purpose we adopt the kurtosis intersection method \cite{Karsch:2001nf}.
The value of kurtosis can be used to diagnose the strength of phase transitions.
For a first order phase transition, the infinite volume value of kurtosis is $K=-2$ while for a crossover it is $K=0$.
At the critical point as the end point of a first order phase transition line, the kurtosis is expected to take the same value irrespective of the spatial volume between $-2$ and $0$.
The value at the critical end point depends on the universality class of the second order phase transition.

Figure~\ref{fig:krtintersection} plots the minimum of kurtosis as a function of $\beta$ for some selected values of $a\mu$.
This shows that a strong first order phase transition at lower $\beta$ becomes weaker for higher $\beta$ and such a change becomes rapid for larger volumes.
We fit the data with the fitting form \cite{deForcrand:2006pv} inspired by finite size scaling,
\begin{equation}
K_{\rm min}
=
K_{\rm E}
+
A N_{\rm s}^{1/\nu}(\beta-\beta_{\rm E}),
\label{eqn:Kintersectionformula}
\end{equation}
where $K_{\rm E}$, $A$, $\nu$ and $\beta_{\rm E}$ are fitting parameters and the results are listed in Table~\ref{tab:kurtosisintersectionfit}.
The resulting exponent $\nu$ and the value of kurtosis at the critical end point $K_{\rm E}$ are independent of $a\mu$ within errors,  and they are consistent with the values of 3-dimensional ${\rm Z}_2$ universality class, $\nu=0.63$ and $K_{\rm E}=-1.396$ respectively.
On the other hand, the universality class of 3-dimensional O(2) and 3-dimensional O(4) are rejected, rather strongly by the value of $K_E$.

We superimpose the obtained critical end points $(\beta_{\rm E}(a\mu),\kappa_{\rm E}(a\mu))$ for $0\le a\mu \le0.19$ in the phase diagram of Figure~\ref{fig:betakappa}.
The critical end point moves toward the upper-left corner by increasing $a\mu$.

\begin{table}[t]
\begin{tabular}{l|llllll}
\hline
\hline
$a\mu$
&
$\beta_{\rm E}$
&
$\kappa_{\rm E}$
&
$K_{\rm E}$
&
$\nu$
&
$A$
&
$\chi^2/{\rm dof}$
\\
\hline
$0.00$&$1.7247 ( 10 )$&$0.1406320 ( 29 )$&$-1.366 ( 15 )$&$0.734 ( 35 )$&$0.64 ( 10 )$&$3.76$\\
$0.05$&$1.72249 ( 92 )$&$0.1406904 ( 29 )$&$-1.383 ( 14 )$&$0.694 ( 32 )$&$0.545 ( 86 )$&$3.24$\\
$0.10$&$1.71803 ( 76 )$&$0.1407928 ( 29 )$&$-1.403 ( 12 )$&$0.615 ( 27 )$&$0.371 ( 64 )$&$3.18$\\
$0.15$&$1.71372 ( 95 )$&$0.1408541 ( 31 )$&$-1.381 ( 15 )$&$0.569 ( 33 )$&$0.281 ( 67 )$&$5.54$\\
$0.16$&$1.7129 ( 11 )$&$0.1408622 ( 33 )$&$-1.371 ( 18 )$&$0.567 ( 38 )$&$0.279 ( 78 )$&$4.94$\\
$0.17$&$1.7119 ( 14 )$&$0.1408723 ( 38 )$&$-1.358 ( 22 )$&$0.570 ( 49 )$&$0.285 ( 99 )$&$3.89$\\
$0.18$&$1.7109 ( 20 )$&$0.1408864 ( 48 )$&$-1.346 ( 28 )$&$0.590 ( 74 )$&$0.33 ( 15 )$&$2.86$\\
$0.19$&$1.7095 ( 32 )$&$0.1409120 ( 62 )$&$-1.335 ( 43 )$&$0.66 ( 13 )$&$0.47 ( 33 )$&$2.03$\\
\hline
\multicolumn{2}{}{}&&&&&\\
\hline
\multicolumn{3}{l|}{Universality class} &$K_{\rm E}$&$\nu$&$\gamma/\nu$&\\
\hline
\multicolumn{3}{l|}{3-dimensional ${\rm Z}_2$} &$-1.396$&$0.63$&$1.964$&\\
\multicolumn{3}{l|}{3-dimensional ${\rm O}(2)$} &$-1.758$&$0.672$&$1.962$&\\
\multicolumn{3}{l|}{3-dimensional ${\rm O}(4)$} &$-1.908$&$0.748$&$1.975$&\\
\hline
\hline
\end{tabular}
\caption{
Fit results for kurtosis intersection for selected values of $a\mu$.
The errors are estimated by the jackknife method.
$\chi^2/{\rm dof}$ is the average value.
In the region $0\le a\mu \le 0.19$, the exponent $\nu$ and the value of kurtosis at the critical end point $K_{\rm E}$ are constant within errors and the universality class is consistent with 3-dimensional ${\rm Z}_2$,
while other universality classes, 3-dimensional O(2) and 3-dimensional O(4) are rejected.
}
\label{tab:kurtosisintersectionfit}
\end{table}

\begin{figure}[t]
\begin{center}
\begin{tabular}{c}
\scalebox{1.}{\includegraphics{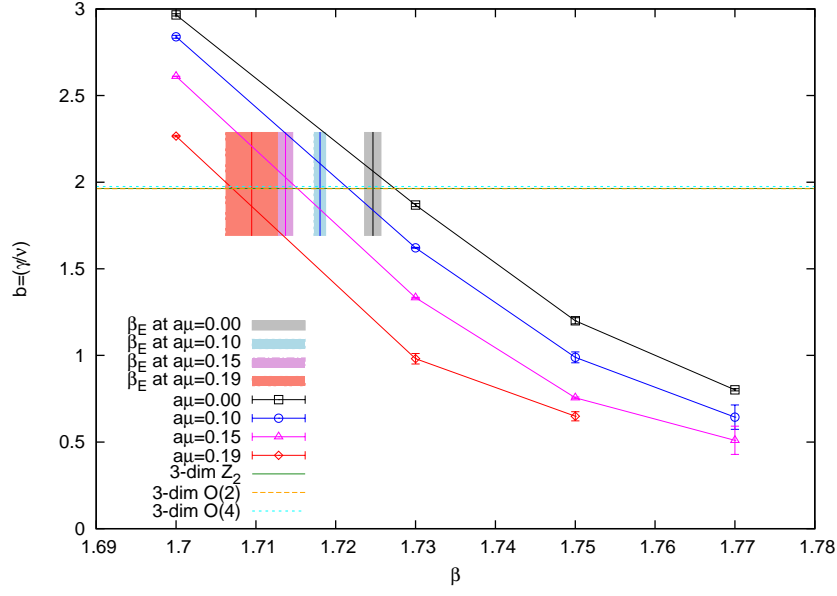}}
\end{tabular}
\caption{
Exponent $b$ of peak height of susceptibility as a function of $\beta$ (open symbols).
The lines connecting points are just for a guide for your eyes.
The filled regions represent the corresponding critical end point determined by the kurtosis intersection method.
The horizontal three lines ($b\approx2$) represent the values of $\gamma/\nu$ for universality classes Z${}_2$, O(2) and O(4) in 3-dimension.
It is hard to see their difference at this scale.
}
\label{fig:exponent}
\end{center}
\end{figure}

In order to confirm the universality class and the location of the critical end point, we check another exponent $\gamma/\nu$ which is obtained from the volume scaling of the susceptibility peak height of quark condensate
\be
\chi_{\rm max}=CN_{\rm s}^b,
\ee
with fit parameters $b$ and $C$.
The exponent $b$ depends on the nature of  transition, {\it i.e.}, $b=d$ the spatial dimensionality at a  first order phase transition, and $b=0$ for a crossover.
At the critical point as the boundary of the first order phase transition line, the exponent is expected to be $b=\gamma/\nu$ with critical exponents $\gamma$ and $\nu$.
Figure~\ref{fig:exponent} shows the exponent $b$ along the transition line as a function of $\beta$.
We observe that the exponent $b$ at the critical end point estimated by the kurtosis intersection
is consistent with the value for the 3-dimensional ${\rm Z}_2$, $\gamma/\nu=1.964$.
Thus we observe a consistency between the kurtosis intersection analysis and the volume scaling of susceptibility.
We note that it is difficult to differentiate universality classes depending solely on the volume scaling of the susceptibility peak since the values of
$\gamma/\nu$ are very close to each other as listed in Table~\ref{tab:kurtosisintersectionfit}.

\subsection{Critical line}
\label{subsec:criticalline}

The analysis of the critical line below requires a careful manipulation with scale setting.  Thus we distinguish quantities in lattice units from those  in physical units by placing a tilde on the former, {\it e.g.,} the chemical potential in physical units is denoted as $\mu$ and that in lattice units by  $\tilde\mu=a\mu$. 

In the previous subsection, we have determined the critical end points in the bare parameter space.
The last step is to translate the critical end point on the $(\beta,\kappa)$ plane to the physical space, to obtain the critical line as one varies $\mu$, and  finally to extract its curvature.
For that purpose, as explained in Sect.~\ref{sec:strategy}, we need to compute the pair of ratios in eq.(\ref{eqn:ratios}) as follows,
\bea
\frac{m_{\rm PS,E}(\mu)}{m_{\rm PS,E}(0)}
&=&
\frac{\tilde m_{\rm PS,E}(\tilde\mu)}{\tilde m_{\rm PS,E}(0)}
\cdot
\frac{a(0)}{a(\tilde\mu)},
\label{eqn:A}
\\
\frac{\mu}{T_{\rm E}(0)}
&=&
\tilde\mu
\cdot
\frac{a(0)}{a(\tilde\mu)}
\cdot
N_{\rm t},
\label{eqn:B}
\eea
where $\tilde m_{\rm PS,E}(\tilde\mu)$ is the pseudo-scalar (PS) meson mass in lattice units evaluated at $(\beta,\kappa)=(\beta_{\rm E}(\tilde\mu),\kappa_{\rm E}(\tilde\mu))$.
Note that the PS mass at the critical point does not depend on $\tilde\mu$ directly, but only through $\beta_{\rm E}$ and $\kappa_{\rm E}$ at $\tilde\mu$.
The PS mass is measured by the zero temperature simulation at $\beta_{\rm E}$ and $\kappa_{\rm E}$.
On the other hand, the lattice spacing requires some careful thought as follows.

We usually determine the lattice spacing by choosing a line of constant physics (LCP) and specifying the value of a dimensionful physical quantity on that line.  For example, one may choose the dimensionless combination $m_{\rm PS}\sqrt{t_0}$ for specifying the LCP, and the value of
$m_{\rm PS}$
in physical units to determine the lattice spacing along the chosen LCP,
\be
a(\beta,y)
=
\frac{\tilde m_{\rm PS}(\beta,\kappa_y(\beta))}{m_{\rm PS}(y)},
\label{eqn:latticespacingLCP}
\ee
where $y$ is the value of the constant physics $y=m_{\rm PS}\sqrt{t_0}$ and $\kappa_y(\beta)$ is defined such that the following equation holds for each $\beta$,
\be
y=\tilde m_{\rm PS}(\beta,\kappa_y(\beta)) \sqrt{\tilde t_0}(\beta,\kappa_y(\beta)).
\ee
The notation of the lattice spacing in eq.(\ref{eqn:A},\ref{eqn:B}) means that
\be
a(\tilde\mu)
=
a(\beta_{\rm E}(\tilde\mu),y).
\ee
Note that, again, the lattice spacing does not depend on $\mu$ directly,
but only though the $\beta_{\rm E}$ at $\tilde\mu$.
Thanks to LCP, where the physical unit mass in the denominator in eq.(\ref{eqn:latticespacingLCP}) is not known a priori but common,
the physical mass cancels out in the ratio of lattice spacings 
and the ratio may be computed by using the PS mass in lattice units,
\be
\frac{a(0)}{a(\tilde\mu)}
=
\frac
{\tilde m_{\rm PS}(\beta_{\rm E}(0),\kappa_y(\beta_{\rm E}(0)))}
{\tilde m_{\rm PS}(\beta_{\rm E}(\tilde\mu),\kappa_y(\beta_{\rm E}(\tilde\mu)))}.
\ee
In the following, for the computation of the ratio of lattice spacings,
we use the Wilson flow scale instead of the PS mass since the former is precisely calculated
\be
\frac{a(0)}{a(\tilde\mu)}
=
\frac
{1/\sqrt{\tilde t_0}(\beta_{\rm E}(0),\kappa_y(\beta_{\rm E}(0)))}
{1/\sqrt{\tilde t_0}(\beta_{\rm E}(\tilde\mu),\kappa_y(\beta_{\rm E}(\tilde\mu)))}.
\ee

One can employ a different LCP by specifying a different value of $y^\prime(\neq y)$.
The resulting lattice spacing coincides with that from the original ($y$) definition if, in specifying the value of the dimensionful quantity, one takes into account  the variation of that quantity in moving from the original LCP to a new LCP.  In general the agreement will not be exact due to scaling violations. 
\be
a(\beta,y^\prime)
=
a(\beta,y)
+
\mbox{(lattice artifacts)}.
\ee
Thus differences one may observe in physical results due to the choice of LCP is a scaling violation effect. 
In the following, we choose two values for the line of constant physics,
\be
y=m_{\rm PS}\sqrt{t_0}=0.55 \mbox{  and  } 0.65.
\ee

\begin{figure}[t]
\scalebox{1.2}{\includegraphics{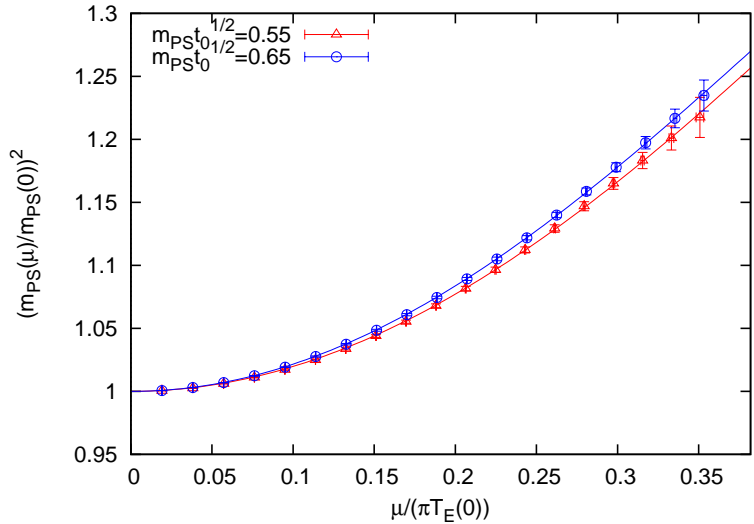}}
\caption{
Critical line for constant physics
$m_{\rm PS}t_0^{1/2}=0.55, 0.65$
with scale input: the Wilson flow scale.
Both lines are bending toward the upper direction.
\label{fig:criticalline}
}
\end{figure}

We use the Wilson flow scale and the hadron mass computed in Appendix A of Ref.~\cite{Jin:2014hea}, where the zero temperature simulations were carried out with the same lattice actions and sufficiently large lattices $m_{\pi}L > 5$.
Especially, we select $\beta=1.70$, $1.73$, $1.75$ and $1.77$ data in our analysis here.
By combining the above scale inputs and the information of the critical end point at finite chemical potential determined in the previous subsection, we calculate the two ratios in eq.(\ref{eqn:A}) and (\ref{eqn:B}).  
The results are plotted in Figure~\ref{fig:criticalline}.


We extract the curvature by using the fitting form in eq.(\ref{eqn:MPSmuT}).  The results are tabulated in Table \ref{tab:curvature}.   
The errors of fitted parameters are estimated by the jackknife method using the uncorrelated chi squared function in each fit.
We also try to perform a fit including correlations by using the covariance matrix estimated by the jackknife method; the results are consistent with the above analysis although the covariance matrix is poorly estimated.
We observe that the critical line has a sensitivity on the value of constant physics.
This difference is considered as a systematic uncertainty caused by the choice of the scale setting as discussed above.  All in all, we find the curvature of the critical line to be positive with a statistical error of about 3\% and a systematic error of about 10\%.

\begin{table}[t]
\begin{center}
\begin{tabular}{l|l|l|l}
\hline\hline
constant physics
&
scale input
&
$\alpha_1$
&
$\alpha_2$
\\
\hline
$m_{\rm PS}\sqrt{t_0}=0.55$
&
$1/\sqrt{t_0}$
&
$1.924 ( 60 )$
&
$-0.58 ( 72 ) $
\\
$m_{\rm PS}\sqrt{t_0}=0.65$
&
$1/\sqrt{t_0}$
&
$2.148 ( 39 )$
&
$-1.74 ( 52 )$
\\
\hline\hline
\end{tabular}
\end{center}
\caption{
The curvature of the critical line in the fitting form in eq.(\ref{eqn:MPSmuT}) for scale inputs ($\sqrt{t_0}$) and the values of constant physics
$m_{\rm PS}\sqrt{t_0}=0.55,0.65$.
The error of curvature is estimated by the jackknife method.
\label{tab:curvature}
}
\end{table}

\section{Concluding remarks}
\label{sec:conclusion}

We have investigated the critical line on the $\mu$-$m_{\pi}$ plane, especially its curvature, in $N_{\rm f}=3$ QCD by using non-perturbatively $O(a)$ improved Wilson fermion action.
We have determined the critical end point by making use of the kurtosis intersection method.  The critical line is drawn by repeating the calculation in the range of chemical potential with applications of various reweighting techniques, that is, the multi-parameter/phase/multi-ensemble reweighting.

The value of kurtosis at the critical end point and the exponent $\nu$ obtained from the kurtosis intersection analysis in the range of chemical potential we investigated are consistent with those of the 3-dimensional ${\rm Z}_2$ universality class.
Furthermore, if the above universality class is used as an input in the analysis of exponent extracted from susceptibility peak, the expected location of the critical point is consistent with that obtained from the kurtosis intersection method.

Our analysis shows that the curvature of the critical line is positive.  This disagrees with a previous study with the naive staggered fermion action \cite{deForcrand:2006pv,deForcrand:2007rq} where the critical line is expressed in terms of quark mass.
We note that neither the previous study nor ours have taken the continuum limit.
Thus further work with larger $N_{\rm t}$ is desired.

\begin{acknowledgments}
BQCD code~\cite{BQCD} was used in this work.
This research used computational resources of the K computer provided by the RIKEN Advanced Institute for Computational Science
through the HPCI System Research project (Project ID:hp120115), and the HA-PACS
provided by Interdisciplinary Computational Science Program
in Center for Computational Sciences, University of Tsukuba.
This work is supported by JSPS KAKENHI 
Grant Numbers 23740177 and 26800130.
This work was supported by FOCUS Establishing Supercomputing Center of Excellence.

\end{acknowledgments}

\appendix

\section{Reweighting}
\label{sec:reweighting}

The phase reweighting and multi-parameter reweighting for bare parameters\footnote{
We use the bare mass parameter $m_0$ instead of $\kappa=1/(2m_0+8)$, since the former parameter is useful in the following discussion.
In this appendix, $\mu$ is the chemical potential in lattice units.
We do not consider $\beta$-reweighting.
}
$m_0$ and $\mu$
can be done by the formula
\be
\langle
{\cal O}(m_0^\prime,\mu^\prime)
\rangle_{m_0^\prime,\mu^\prime}
=
\frac{
\left\langle
\left(
\frac{\det D(m_0^\prime,\mu^\prime)}{\det D(m_0,\mu)}
\right)^{N_{\rm f}}
e^{iN_{\rm f}\theta(m_0,\mu)}
{\cal O}(m_0^\prime,\mu^\prime)
\right\rangle_{||,m_0,\mu}
}
{
\left\langle
\left(
\frac{\det D(m_0^\prime,\mu^\prime)}{\det D(m_0,\mu)}
\right)^{N_{\rm f}}
e^{iN_{\rm f}\theta(m_0,\mu)}
\right\rangle_{||,m_0,\mu}
}.
\label{eqn:phaseparameterreweighting}
\ee
Here, $m_0^\prime$ and $\mu^\prime$ are target parameters
while $m_0$ and $\mu$ are actual simulation parameters.
The average $\langle...\rangle_{m_0^\prime,\mu^\prime}$ in LHS is taken by using the Boltzmann factor including the full quark determinant at parameter $m_0^\prime$, $\mu^\prime$,
while the average $\langle...\rangle_{||,m_0,\mu}$ in RHS is taken by using the Boltzmann factor including the phase quenched quark determinant at parameter $m_0$, $\mu$.
Here we have explicitly written down the bare parameter dependence on the observable ${\cal O}(m_0,\mu)$, say the quark propagator.

In eq.(\ref{eqn:phaseparameterreweighting}), one needs to evaluate the reweighting factor,
\be
\left(
\frac{\det D(m_0^\prime,\mu^\prime)}{\det D(m_0,\mu)}
\right)^{N_{\rm f}}
e^{iN_{\rm f}\theta(m_0,\mu)},
\ee
where the second factor is already computed and stored but the first factor, the ratio of quark determinants, requires high cost computation if one tries to calculate it directly at many target parameter points $(m_0^\prime,\mu^\prime)$.
Thus we adopt a cheaper approximation method, that is, the Taylor expansion of the logarithm of determinant which is known to have better convergence property than the other expansion schemes \cite{Jin:2013wta},
\be
\ln
\left(
\frac{\det D(m_0^\prime,\mu^\prime)}{\det D(m_0,\mu)}
\right)
=
\left[
\sum_{j,k=0}^\infty
\frac{\Delta m_0^j \Delta\mu^k}{j!k!}
\left(
\frac{\partial}{\partial m_0}
\right)^j
\left(
\frac{\partial}{\partial \mu}
\right)^k
\ln \det D(m_0,\mu)
\right]
-
\ln \det D(m_0,\mu),
\label{eqn:ratiodeterminantexpansion}
\ee
with
\bea
\Delta m_0&=&m_0^\prime-m_0,\\
\Delta\mu&=&\mu^\prime-\mu.
\eea
Once some leading coefficients in the expansion are calculated, one can easily evaluate the ratio at many reweighted points up to truncation errors.
In our calculation, we include the following coefficients
\bea
(j,k)
=
\left\{
\begin{array}{ccl}
(1-4,0)\hspace{5mm}&:&\mbox{purely }m_0\mbox{-derivatives},\\
(0,1-4)\hspace{5mm}&:&\mbox{purely }\mu\mbox{-derivatives},
\end{array}
\right.
\eea
but no mixed derivatives say $(1,1)$ and so on.
The explicit form of the approximated ratio of determinant is given by
\be
\left(
\frac{\det D(m_0^\prime,\mu^\prime)}{\det D(m_0,\mu)}
\right)^{N_{\rm f}}
\approx
\exp\left[
N_{\rm f}
\sum_{j=1}^4
\frac{(-)^{j+1}\Delta m_0^j}{j}
{\rm Tr}
D^{-j}(m_0,\mu)
+
N_{\rm f}
\sum_{k=1}^4
\frac{(\Delta\mu/T)^k}{k!}
W_k(m_0,\mu/T)
\right],
\label{eqn:ratiodetapproximation}
\ee
where $W_{1,2,3,4}$ are given in \cite{Jin:2013wta}.

For the same reason as the ratio of determinant,
we use an expansion form of the observable in eq.(\ref{eqn:phaseparameterreweighting})
\be
{\cal O}(m_0^\prime,\mu^\prime)
=
\sum_{j,k=0}^\infty
\frac{\Delta m_0^j \Delta\mu^k}{j!k!}
\left(
\frac{\partial}{\partial m_0}
\right)^j
\left(
\frac{\partial}{\partial \mu}
\right)^k
{\cal O}(m_0,\mu).
\ee
For the trace of higher powers of quark propagator which is included in the higher moments of quark condensate\footnote{The formulae for higher moments of quark condensate in terms of quark propagator are explicitly given in Ref.~\cite{Jin:2014hea}.},
we apply the following approximation ($\mu$-derivative terms are totally neglected)
\bea
{\rm tr} D^{-1}(m_0^\prime,\mu^\prime)
&\approx&
{\rm tr} D^{-1}(m_0,\mu)
+\sum_{j=1}^3
(-)^j\Delta m_0^j
{\rm tr} D^{-(j+1)}(m_0,\mu),
\label{eqn:D1}
\\
{\rm tr} D^{-2}(m_0^\prime,\mu^\prime)
&\approx&
{\rm tr} D^{-2}(m_0,\mu)
+\sum_{j=1}^2
(-)^j(j+1)\Delta m_0^j
{\rm tr} D^{-(j+2)}(m_0,\mu),
\label{eqn:D2}
\\
{\rm tr} D^{-3}(m_0^\prime,\mu^\prime)
&\approx&
{\rm tr} D^{-3}(m_0,\mu)
+\sum_{j=1}^1
(-)^j(j+1)(j+2)\Delta m_0^j
{\rm tr} D^{-(j+3)}(m_0,\mu),
\label{eqn:D3}
\\
{\rm tr} D^{-4}(m_0^\prime,\mu^\prime)
&\approx&
{\rm tr} D^{-4}(m_0,\mu).
\label{eqn:D4}
\eea

At first glance, you may doubt the approximation for the ratio of determinant in eq.(\ref{eqn:ratiodetapproximation}) and observables in eq.(\ref{eqn:D1}-\ref{eqn:D4}) especially the higher powers of the inverse.
We however have some evidences that this approximation is good within our parameter range and statistical error.
First, the approximation for the observables in eq.(\ref{eqn:D1}-\ref{eqn:D4}) is compared with the partial quenching results where there is no truncation error in the observable.
Even for the kurtosis including eq.(\ref{eqn:D4}), we do not see any significant difference between them within errors.
Second, in order to check the effects of the mixed derivative terms in the reweighting factor,
we compute and include the mixed-derivative coefficients up to 4th order,
namely $(j,k)=(1,1)$, $(2,1)$, $(3,1)$, $(1,2)$, $(2,2)$, $(1,3)$ coefficients in eq.(\ref{eqn:ratiodeterminantexpansion}),
and check their effects on the moments of chiral condensate (of course,
the associated mixed derivative contributions for the observable are also included)
at $\beta=1.73$ and $N_{\rm s}=12$ in the range of the chemical potential, $0 \le \mu \le 0.19$.
Then it turns out that the difference is quite small, that is,
within statistical errors in the parameter space.
Furthermore, we check the hierarchy of the terms and observe that the dominant term is (1,0)
and the leading term in the mixed derivative terms is (1,1),
and then it turns out that the magnitude of their ratio, $|[(1,1) \mbox{ term}]/[(1,0) \mbox{ term}]|$ 
is of order $10^{-3}$ at a maximum.
This shows that neglecting the mixed derivative terms is justified and our approximation in eq.(\ref{eqn:ratiodetapproximation}) is fine.
We naturally expect that the same goes for the other cases of $\beta=1.70,1.75,1.77$.
Thus, we conclude that the approximation made in eq.(\ref{eqn:ratiodetapproximation}) is legitimate in the range of parameter we explored.

\section{Multi-ensemble reweighting}
\label{sec:multiensemblereweighting}
In this appendix, we review the multi-ensemble reweighting technique
in \cite{reweighting}.
An estimated (E) expectation value
of some operator at a target parameter denoted by T is given by
\be
\Omega_{\rm T}^{\rm E}
=
\frac{
\sum_{U}
w_{\rm T}(U)\Omega_{\rm T}(U)
}{
\sum_{U^\prime}
w_{\rm T}(U^\prime)
},
\ee
where $\sum_{U}$ is an abbreviation of sum over all configurations,
namely sum over all ensemble (each ensemble is numbered by $r$)
and all configurations (numbered by $n$) therein,
\be
\sum_{U}f(U)
=
\sum_{r=1}^{R}\sum_{n=1}^{N_r}
f(U_{r,n})
,
\ee
with the number of ensembles $R$ and the total number of configurations $N_r$
for ensemble $r$.

The reweighting factor $w_{\rm T}$ is given by
\be
w_{\rm T}(U)
=
\frac
{
\sum_{s^\prime=1}^R
N_{s^\prime}
}
{
\sum_{s=1}^R
N_s
\exp
\left[
S_{\rm T}(U)
-
S_s^{||}(U)
+
f_s
-
F_{\rm T}
\right]
},
\label{eqn:weight}
\ee
where
$S_{\rm T}$ is the action at the target parameter,
and $S_s^{||}$ is the simulated actions (using phase quenched determinant at the simulated parameter)
for ensemble $s$ ($s=1,2,...,R$).
In our case\footnote{In contrast to the previous appendix, we use $\kappa$ instead of the bare mass $m_0$ in this appendix. The arguments of the Dirac operator are $\kappa$, $\mu$ and gauge configuration of $U$. Note that $r$ denoting the ensemble of $U_{r,n}$ and $s$ denoting the ensemble of parameter of action $S^{||}_{s}$ are independent.} 
\bea
e^{-S_{\rm T}(U_{r,n})}
&\longrightarrow&
e^{-S_{\rm G}(U_{r,n})}
\det D(\kappa_{\rm T},\mu_{\rm T};U_{r,n}),
\\
e^{-S_s^{||}(U_{r,n})}
&\longrightarrow&
e^{-S_{\rm G}(U_{r,n})}
|\det D(\kappa_s,\mu_s;U_{r,n})|.
\eea
The ratio of Boltzmann weight in eq.(\ref{eqn:weight}) is given by
\bea
\exp
\left[
S_{\rm T}(U_{r,n})
-
S_s^{||}(U_{r,n})
\right]
&\longrightarrow&
\frac{
|\det D(\kappa_s,\mu_s;U_{r,n})|
}{
\det D(\kappa_{\rm T},\mu_{\rm T};U_{r,n})
}
\non\\
&=&
\frac{
\frac{
|\det D(\kappa_s,\mu_s;U_{r,n})|
}
{
|\det D(\kappa_r,\mu_r;U_{r,n})|
}
e^{-i\theta(\kappa_r,\mu_r;U_{r,n})}
}{
\frac{
\det D(\kappa_{\rm T},\mu_{\rm T};U_{r,n})
}{
\det D(\kappa_r,\mu_r;U_{r,n})
}
},
\eea
where we have already measured the phase $\theta(\kappa_r,\mu_r,U_{r,n})$.
The ratio of determinants can be estimated by using the expansion method given
in the previous appendix.

$f_s$ ($s=1,2,...,R$) in eq.(\ref{eqn:weight})
which are free parameter and we determine them by solving the non-linear equation,
\be
f_s
=
F_s
-\ln
\sum_U
\left(
\sum_{s^\prime=1}^R
N_{s^\prime}
\exp
\left[
S_s^{||}(U)
-
S_{s^\prime}^{||}(U)
+
f_{s^\prime}
-
F_s
\right]
\right)^{-1}
\in \mathbb{R},
\ee
where $F_s$ are dummy to avoid numerical instability.
We solve the equation by iteratively substituting trial values of $f_s$
with initial values $f_s=0$ for all $s$.
We observe that this iteration converges after around (or less than) 20 iterations
for all cases.

$F_{\rm T}$ in eq.(\ref{eqn:weight}) is a constant to avoid numerical instability,
\be
F_{\rm T}
=
\frac{1}{\sum_{r=1}^R N_r}
\sum_U
\frac{1}{R}
\sum_{s=1}^{R}
\left[
S_{\rm T}^{||}(U)-
S_{s}^{||}(U)+f_s
\right] \in \mathbb{R}.
\ee


\begin{thebibliography}{99}

\bibitem{Endrodi:2007gc} 
  G.~Endrodi, Z.~Fodor, S.~D.~Katz and K.~K.~Szabo,
  PoS LAT {\bf 2007}, 182 (2007)
  [arXiv:0710.0998 [hep-lat]].
  
\bibitem{Ding:2011du} 
  H.-T.~Ding, A.~Bazavov, P.~Hegde, F.~Karsch, S.~Mukherjee and P.~Petreczky,
  PoS LATTICE {\bf 2011}, 191 (2011)
  [arXiv:1111.0185 [hep-lat]].
  
\bibitem{Kayaetal1999}
JLQCD Collaboration (S. Aoki (Tsukuba U.) et al.), 
Nucl. Phys. Proc. Suppl. {\bf 73},  459 (1999).

\bibitem{Karsch:2001nf} 
  F.~Karsch, E.~Laermann and C.~Schmidt,
  Phys.\ Lett.\ B {\bf 520}, 41 (2001)
  [hep-lat/0107020].
    
\bibitem{Smith11}
  D.~Smith and C.~Schmidt,
  PoS(Lattice 2011), 216 (2011).
  [arXiv:1109.6729[hep-lat]]
 
\bibitem{deForcrand:2006pv} 
  P.~de Forcrand and O.~Philipsen,
  JHEP {\bf 0701}, 077 (2007)
  [hep-lat/0607017].


\bibitem{Jin:2014hea} 
  X.~Y.~Jin, Y.~Kuramashi, Y.~Nakamura, S.~Takeda and A.~Ukawa,
  Phys.\ Rev.\ D {\bf 91}, no. 1, 014508 (2015)
  [arXiv:1411.7461 [hep-lat]].
  

\bibitem{Fodor:2001au} 
  Z.~Fodor and S.~D.~Katz,
  Phys.\ Lett.\ B {\bf 534}, 87 (2002)
  [hep-lat/0104001].
  
\bibitem{Fodor:2001pe} 
  Z.~Fodor and S.~D.~Katz,
  JHEP {\bf 0203}, 014 (2002)
  [hep-lat/0106002].

\bibitem{deForcrand:2010ys} 
  P.~de Forcrand,
  PoS LAT {\bf 2009}, 010 (2009)
  [arXiv:1005.0539 [hep-lat]].

\bibitem{Philipsen:2011zx} 
  O.~Philipsen,
  Acta Phys.\ Polon.\ Supp.\  {\bf 5}, 825 (2012)
  [arXiv:1111.5370 [hep-ph]].
  

\bibitem{deForcrand:2007rq} 
  P.~de Forcrand, S.~Kim and O.~Philipsen,
  PoS LAT {\bf 2007}, 178 (2007)
  [arXiv:0711.0262 [hep-lat]].
    
\bibitem{wilflow}
  M.~L\"uscher,
  JHEP \textbf{1008}, 071 (2010),
  [ arXiv:1006.4518 [hep-lat]].

\bibitem{CPPACS2006}
CP-PACS and JLQCD Collaborations (S. Aoki {\it et al.}), Phys. Rev. D{\bf 73}, 034501 (2006)
[hep-lat/0508031].

\bibitem{iwasaki}
  Y.~Iwasaki, Report No. UTHEP-118 (1983), 
  [arXiv.1111.7054].


\bibitem{RHMC}
  M.~A.~Clark and A.~D.~Kennedy, Phys. Rev. Lett. \textbf{98}, 051601 (2007),
  [arXiv:hep-lat/0608015].
  

\bibitem{Danzer:2008xs} 
  J.~Danzer and C.~Gattringer,
  Phys.\ Rev.\ D {\bf 78}, 114506 (2008)
  [arXiv:0809.2736 [hep-lat]].

\bibitem{Takeda:2011vd} 
  S.~Takeda, Y.~Kuramashi and A.~Ukawa,
  Phys.\ Rev.\ D {\bf 85}, 096008 (2012)
  [arXiv:1111.6363 [hep-lat]].

\bibitem{takedanote} 
  S.~Takeda, Y.~Kuramashi and Y.~Nakamura
  AICS Technical Report No. 2015-001.

  
\bibitem{reweighting}
  A.~M.~Ferrenberg and R.~H.~Swendsen, 
  Phys. Rev. Lett. {\bf 61}, 2635 (1988). 



\bibitem{BQCD}
  Y.~Nakamura and H.~St\"uben,
  PoS(Lattice 2010), 040 (2010),
  [arXiv:1011.0199 [hep-lat]].


\bibitem{Jin:2013wta}
  X.~Y.~Jin, Y.~Kuramashi, Y.~Nakamura, S.~Takeda and A.~Ukawa,
  Phys.\ Rev.\ D {\bf 88} (2013) 9,  094508
  [arXiv:1307.7205 [hep-lat]].

\end{thebibliography}
\end{document}